\documentclass[12pt]{article}
\usepackage[dvips]{graphicx}
\let\jnfont =\rm
\def\NPB#1,{{\jnfont Nucl.\ Phys.\ B }{\bf #1},}
\def\PLB#1,{{\jnfont Phys.\ Lett.\ B }{\bf #1},}
\def\PRD#1,{{\jnfont Phys.\ Rev.\ D } {\bf #1},}
\def\PRL#1,{{\jnfont Phys.\ Rev.\ Lett.\ }{\bf #1},}

\def\ZPC#1,{{\jnfont Z.~Phys.\ C }{\bf #1},}
\begin{document}
\begin{flushright}
VLBL Study Group-H2B-4\\
AMES-HET-01-09 \\
AS-ITP-2001-022\\
\today 
\end{flushright}
\vspace{0.2in}
\begin{center}
{\Large Probing neutrino oscillations jointly\\ in long and very long baseline 
             experiments}
\vspace{.3in}

Y. F. Wang$^a$, K. Whisnant$^b$,  Zhaohua Xiong$^c$, Jin Min Yang$^c$, 
Bing-Lin Young$^b$ 
\vspace{.3in}

{\small \it
$^a$  Institute of High Energy Physics, Academia Sinica, Beijing 100039, China \\
$^b$ Department of Physics and Astronomy, Iowa State University,
     Ames, Iowa 50011, USA \\
$^c$ Institute of Theoretical Physics, Academia Sinica, Beijing 100080, China}
\end{center}
\vspace{.5in}
     \begin{center} ABSTRACT  \end{center}
We examine the prospects of making a joint analysis of neutrino oscillation 
at two baselines with neutrino superbeams.  Assuming narrow band superbeams 
and a 100 kt water Cerenkov calorimeter, we calculate the event rates and 
sensitivities to the matter effect, the signs of the neutrino mass differences, the
CP phase and the mixing angle $\theta_{13}$.  Taking into account all possible 
experimental errors under general consideration, we explored the optimum cases 
of narrow band beam to measure the matter effect and the CP violation effect at 
all baselines up to 3000 km.  We then focus on two specific baselines, a long 
baseline of 300 km and a very long baseline of 2100 km, and analyze their joint 
capabilities.  We found that the joint analysis can offer extra leverage to resolve 
some of the ambiguities that are associated with the measurement at a single baseline.

\section{Introduction}
\label{sec1}
Although the existing data from the Super-Kamiokande experiment~\cite{superK} 
and various other corroborating experiments offer very strong indications of 
neutrino oscillations, the appearance experiment, i.e., the appearance of a flavor 
different from the original one, has not been convincingly performed. 
If neutrinos indeed oscillate,  the oscillation parameters, including the leptonic
CP phase, have to be determined with sufficient accuracy. Furthermore,  the 
well-known MSW matter effect \cite{MSW} has to be tested by experiments.  
In spite of the various ongoing and planed neutrino oscillation experiments, 
additional experiments with very long baseline are needed, at least for the test 
of the matter effect.   The recently approved superbeam facility~\cite{HIPA}, 
which will be available towards the later part of this decade, offers the possibility 
of a very long baseline (VLBL) experiment which, in conjunction with other oscillation 
experiments, can test thoroughly all properties of neutrino oscillations.

Among all neutrino oscillation experiments,  the long baseline (LBL) 
experiments are particularly attractive.  Since the neutrino beams are 
produced in an accelerator according to definite physics criteria with
the detector site chosen accordingly, the experiment can be conducted 
in a more controlled fashion to maximize the physics output. 
Hence the LBL experiments will allow us to make detailed analyses of the 
oscillation parameters so as to provide a complete picture of the physics 
of neutrino oscillation.  As one example of such experiments, a project 
called H2B  is under discussion~\cite{H2B, Japanesegroup, FoM}.  The 
neutrino super-beam for H2B would be from the newly approved high 
intensity 50 GeV proton synchrotron in Japan called HIPA~\cite{HIPA} 
and the detector, tentatively called the Beijing Astrophysics and Neutrino 
Detector (BAND), is envisioned to be a 100 kt water Cerenkov calorimeter
(WCC) with resistive plate chambers (RPC)~\cite{wang} located in Beijing, 
China.  The distance from HIPA to Beijing is about 2100 km. Such a very 
long baseline experiment would be complementary  to the recently proposed 
J2K experiment~\cite{J2K} which will also use the neutrino beam from HIPA 
but with the Super-Kamiokande detector or its update.  The distance from HIPA 
to Super-Kamiokande is about 300 km. 

In this article, we will examine the prospects of investigating neutrino 
oscillations at H2B in conjunction with J2K so that the joint data at the
two widely different baselines can be used in a complementary way to 
provide strong leverage to eliminate some of the ambiguities in the 
determination of oscillation parameters. 
The joint analysis can expand the capability of the parameter search that 
are not attainable by either of the experiments alone. The two baselines 
can work at their respective favorable energy ranges.  The present work 
is to demonstrate this possibility.   But we have not search for the best 
narrow beam energies for the two baselines.  Assuming a narrow band 
meson beam and the above mentioned 100~kt WCC with RPC, we 
simulate the event rates for 5-year operation.  The sensitivity 
of the event rates for the various oscillation parameters will be explored.  
The present work can be regarded partly as a continuation of the study of 
H2B Refs.~\cite{H2B, Japanesegroup,  FoM} and an initial exploration of
the idea of joint analyses of two detectors which we think is appropriate for 
oscillation physics.  In Sec.~\ref{sec2}, we discuss some of the fundamentals 
of neutrino oscillation and LBL experiments.  In Sec.~\ref{sec3}, we present 
some of our numerical results.  We present the joint analyses of the data of
two detectors in Sec.~4.  Finally, in Sec.~5, we present our conclusions.
 
\section{Fundamentals of neutrino oscillation and LBL experiments}
\label{sec2}

If we accept all current data, there will be three distinctive mass scales provided  
by the five categories of experiments: long baseline, short baseline accelerator experiments
such as LSND, atmospheric, solar, and reactor.  If the LSND data are excluded, 
the three SM neutrino flavors are sufficient and no extension of the number of 
neutrinos beyond that of the standard model is  necessary.  In view of the 
uncertainty of the LSND data, our discussion will  be restricted to the 3-flavor 
scenario. 

The oscillation of  the 3-flavor neutrinos is a system with a limited number 
of degrees of freedom. The system consists of 2 mass square differences 
(MSD), three mixing angles and one measurable CP phase.  These parameters 
together with the matter effect determine the various survival and appearance 
probabilities~\cite{BDWY}.  The unitary mixing matrix in vacuum is 
generally parameterized as
\begin{eqnarray} 
U & = & \left( \begin{array}{ccc}
    c_{12}c_{13}  & c_{13}s_{12}  &  \hat{s}^*_{13} \\
    -c_{23}s_{12} - c_{12}\hat{s}_{13}s_{23} &
      c_{12}c_{23} -s_{12}\hat{s}_{13}s_{23} & c_{13}s_{23} \\
    s_{12}s_{23} - c_{12}c_{23}\hat{s}_{13} &
    -c_{12}s_{23} -c_{23}s_{12}\hat{s}_{13}  & c_{13}c_{23}
    \end{array} \right)
\end{eqnarray}
where $s_{jk}=\sin(\theta_{jk})$, $c_{jk}=\cos(\theta_{jk})$, and
$\hat{s}_{jk}=\sin(\theta_{jk})e^{i\delta}$, $\theta_{jk}$ defined 
for $j<k$ is the mixing angle of mass eigenstates $\nu_j$ and  $\nu_k$,
and $\delta$ is the CP phase angle.  The three mass eigenvalues are
denoted as $m_1$, $m_2$, and $m_3$.   The two independent MSD
are $\Delta{\rm m}^2_{21}\equiv {\rm m}^2_2 - {\rm m}^2_1$ and
$\Delta{\rm m}^2_{32}\equiv {\rm m}^2_3 - {\rm m}^2_2$.  

In LBL experiments the neutrino beam has to go through matter which 
gives rise to the well-known MSW effect \cite{MSW}.
A widely used model for the Earth, called the preliminary reference Earth model 
PREM, is given in \cite{earth} and the earth density profile can be found in  
\cite{profile}.  Since for a VLBL experiment the matter density can vary 
 significantly along the path of the neutrino beam, in our calculation we perform 
numerical integration of the Schr\"{o}dinger equation for a realistic treatment of 
the  distance dependent matter density. 

The detection of a given neutrino flavor is through its accompanying charged  
lepton produced by the charge current interaction of the neutrino with the 
nucleons in the detector mass. For a neutrino energy $E_\nu$, which is small  
compared to the mass of the W and Z bosons but large enough so that 
quasi-elastic effect is small, the charge current cross sections are given by  
$\sigma_{\nu N}  =  0.67\times 10^{-38}{\rm cm}^2 E_\nu(\rm GeV)$  
for electron and muon neutrinos, and 
$\sigma_{\bar{\nu} N}  =  0.34\times 10^{-38}{\rm cm}^2 E_\nu(\rm GeV)$  
for electron and muon anti-neutrinos. 
For the tau neutrino, the above expression is subject to a threshold suppression.   
The threshold for the production of the tau is
$E_T = m_\tau + {m_\tau^2\over 2m_N}= 3.46~{\rm GeV}$. 
A fit of $\nu_\tau$ to $\nu_\mu$ cross section as a function of the neutrino
energy in terms of the ratio of two quadratic polynomials can be found in 
Ref.~\cite{H2B}.  The signal events of flavor $\beta$, i.e., the number of 
charged lepton of flavor $\beta$, from a neutrino beam of flavor $\alpha$, 
to be observed at a baseline L is given by
\begin{eqnarray}
N_s = \int^{E_{\rm Max}}_{E_{\rm min}} \Phi(E_\nu,L)\sigma(E_\nu) 
   P_{\alpha\to\beta}(E_\nu, L) d E_\nu, 
\end{eqnarray}
where $\Phi(E_\nu,L)$ is the total neutrino flux spectrum including the detector 
size and running time period, $P_{\alpha\to\beta}(E_\nu, L)$ is the oscillation 
probability, $\sigma(E_\nu)$ the neutrino charge current cross section, and 
$E_{\rm Max}$ and $E_{\rm min}$ are the maximum and the minimum 
energies of the beam.  

In a narrow band beam the neutrino flux is distributed below a given energy 
$E_{\rm peak}$. The intensity is peaked at  $E_{\rm peak}$ and decreases
rapidly below $E_{\rm peak}$.  The wide 
band beam contains neutrinos with energy spread out in a significant range 
of energy. In our calculation we will use the realistic beam energies and
profiles provided in \cite{Japanesegroup,beam}.  Some of the narrow band
beams together with the wide beam are plotted in Fig.~1.
Here $dN_{cc}/dE_{\nu} \equiv \Phi(E_\nu, L)\sigma (E_\nu)$ is the 
energy distribution of the charged-current events $N_{cc}$ 
for one year operation of a 100 kt detector at L=2100 km.
    
Since in oscillation experiments, especially in the case of electron neutrino 
appearance, the statistics are generally not large.  Therefore the error is an
important factor in the physics extraction.  W use the approach of
Ref.~\cite{FoM} to estimate the possible statistical and systematic errors
and to gain a sense of the goodness of the fit.  For the electron counting 
experiments the errors and uncertainties arise from the following sources:
\begin{itemize}
{\setlength\itemsep{-0.6ex}
\item[{\rm(i)}] The statistical error in the measurement of the charge lepton of 
                flavor $\beta$ which is as usual $\sqrt{N_s+N_b}$.  
                $N_b$ is the number of measured background events and can be 
                expressed as 
                \begin{eqnarray}
       N_b=f_\beta  \int^{E_{\rm Max}}_{E_{\rm min}} 
                                             \Phi(E_\nu,L)\sigma(E_\nu) d E_\nu .
                \end{eqnarray}
\item[{\rm(ii)}] The systematic uncertainty in the calculation of the number of 
                 background events, which can be denoted as $r_\beta N_b$.  
\item[{\rm(iii)}] The systematic uncertainty in the beam flux and the cross section 
                  which we denote as $g_\beta N_s$.  
}
\end{itemize}
The total error is the quadrature of all these 
uncertainties. In our calculation we will take
$r_\beta=0.1$, $g_\beta=0.05$, and $f_{\beta}=0.01$.  

\section{Numerical results for individual baselines}
\label{sec3}
Presently there are sizable errors in all the oscillation parameters.  However, we 
envisage that at the H2B time, $\Delta{\rm m}^2_{32}$, $\Delta{\rm m}^2_{21}$, 
$\theta_{23}$, and $\theta_{12}$ will be fairly accurately determined.  So we will
not assign any specific errors to them. We focus our investigations on the following 
parameters and effects: matter, MSD sign, CP violation, and $\theta_{13}$. 

\subsection{Inputs}
We present numerical results of a 5-year operation with a water Cerenkov 
detector. The detector size is assumed to be 100 kt for all baselines.  Sizes 
other than 100 kt will be labeled whenever used. 

The inputs of the mixing angles and MSD's are from solar, atmospheric and 
CHOOZ experiments.  For definiteness we take 
$\sin^2(2\theta_{12})=0.8$ and $\sin^2(2\theta_{23})=1.0$.
In most of our results we use $\sin^2(2\theta_{13})=0.05$ for illustration 
and effects of larger and smaller values of $\theta_{13}$, 
$0.01 \leq \sin^2(2\theta_{13})\leq 0.1$, will be investigated.  The inputs 
of MSD $\Delta{\rm m}^2_{21}$ and $\Delta{\rm m}^2_{32}$ are 
respectively given by 
$\Delta{\rm m}^2_{\rm sol}=5 \times 10^{-5}$ eV$^2$  and 
$\Delta{\rm m}^2_{\rm atm}=3 \times 10^{-3}$ eV$^2$.  
 
Presently the sign of the MSD's are unknown so there are 4 possibilities: 
\begin{equation}
\begin{tabular}{lllll} \\ \hline\hline
    & ~~~I & ~~~II & ~~~III & ~~~IV   \\   \hline 
$\Delta{\rm m}^2_{32}$ &  ~~~+ & ~~~+ &  ~~~~- & ~~~~-    \\  \hline
$\Delta{\rm m}^2_{21}$ &  ~~~+ & ~~~ - & ~~~+ & ~~~~-    \\  
\hline\hline
\end{tabular}
\end{equation}

 \vskip 1.ex 
\noindent
After showing the effects of all four sign combinations in the electron event
numbers we will choose the sign I for illustration.  

\subsection{Matter effects}

In Tables 1 and 2 we show the $\nu_\mu\rightarrow \nu_e$ event rates 
with and without matter effects for a narrow band beam with 
$E_{\rm peak}=4$ GeV for both baselines.  It is clear that for both narrow 
band and wide band beams the matter effect is significant on electron event 
number at L=2100 km, but negligible at L=300 km. As expected, 
the $\nu_\mu$ and $\nu_\tau$ events show very little matter effect at either 
distance.    
The event rates at both baselines can be increased if different narrow band 
beams are used.  For example, for L=2100 km the $E_{\rm peak}$=6 GeV 
beam has twice as many electron events as the $E_{\rm peak}=4$ GeV beam.

In order to look for the optimum beam energy to measure matter effects at a 
given baseline, we have examined the following ratio, which is approximately 
the statistical significance of the matter effect and is referred to in 
Ref.~\cite{FoM} as the figure of merit, 
\begin{equation}
R_{\rm matter}=\frac{N_e\vert_{\rm with~ matter} - N_e\vert_{\rm without ~matter}}
                    { \Delta N_e} .
\end{equation}
Here $\Delta N_e$ is the total error of the electron event number, as discussed 
at the end of Sec. 2,  without the matter effect.  Figure 2 shows 
$R_{\rm matter}$ versus the baseline up to 3000 km for several narrow band 
beams for the four MSD signs combinations.  We see that for L=2100 km the 
optimal narrow band beams for the matter effect are with peak energies in the 
range of $4\sim 6$ GeV.  For example, as shown in Fig.~2 for the MSD sign I, 
the optimal narrow band beam has the peak energy around $E_{\rm peak}=4$ 
GeV.  For L=300 km, as expected, there is very little statistical sensitivity to 
the matter effect at all available energies.
 
Given a narrow band beam with $E_{\rm peak}=4$ GeV for  L=2100 km and
 $E_{\rm peak}=0.7$ GeV for  L=300 km, Fig.~3 shows the electron event 
rate versus the CP phase with or without matter effect.  We see that for
$\theta_{13}$ to have a fixed value or small range of uncertainties the matter 
effect is experimentally measurable for L=2100 km but hardly observable for 
L=300 km.  However in the currently fully allowed range of $\theta_{13}$,
$\sin^2(2\theta_{13})\leq 0.1$, it is even difficult for the 2100 km baseline 
to distinguish the matter effect from the vacuum for the following fact:   
Since the electron event rate is proportional to $\sin^2(2\theta_{13})$, 
the electron event rates for $\sin^2(2\theta_{13})=0.03$ with matter effect 
and for $\sin^2(2\theta_{13})=0.1$ in the case of vacuum are the same
as can be inferred from Fig.~3, it is not possible to distinguish the two.
This ambiguity will be reinforced when the error is not negligible. 

\subsection{MSD sign effects} 

The sensitivity of the event rate to the sign of MSD for 
$\sin^2(2\theta_{13})=0.05$ is also shown in
Tables 1 and 2 for $E_{\rm peak}=4$ GeV and $\delta=0$ for both 
baselines, and in Fig.~4 for different energies for the two baselines as
functions of the CP phase.  Tables 1 and 2 show that the electron 
event rates are sensitive to the sign of MSD at the 2100 km baseline.  
It is also interesting to note that for L=300 km  
there is sensitivity in distinguishing signs I and IV in which both 
MSD are positive or negative from signs II and III in which one is 
positive and the other negative.  This general feature is valid for other
values of $\theta_{13}$ once it is determined. 

In Fig.~4, in which we take $\sin^2(2\theta_{13})=0.05$, it shows clearly that 
for L=2100 km I and II are well separated from III and IV for all values of 
CP phase.  Hence the sign of $\Delta{\rm m}^2_{32}$ should be readily 
determined with moderate amount of electron neutrino appearance data.  
However, the separation of I from II depends on the value of the CP phase.  
In the region of small, intermediate and large value of the CP phase, the sign 
of $\Delta{\rm m}^2_{\rm sol}$ can be determined, but 
around $\delta = 130^\circ$ and $\delta = 280^\circ$ I and II are not 
distinguishable.  The signs III and IV are almost inseparable in the whole 
region of $\delta$.  Hence the sign of $\Delta{\rm m}^2_{21}$ will be very 
hard to determine if $\Delta{m}^2_{32} < 0$.  Then the anti-neutrino beam  
is needed for the determination.  For L=300 km, Fig.~4 shows that it is
difficult to distinguish I, II, III and IV except in very special values of 
the CP phase. 
 
Unfortunately, the above result is only true if $\theta_{13}$ is already known.  
Similar to the situation discussed at the end of the preceding subsection, the 
significant uncertainty in $\sin^2(2\theta_{13})$ muddies the water.  As 
$\sin^2(2\theta_{13})$ decreases the electron event rate will also be reduced.
Therefore, it is difficult to distinguish the signs I and II of small $\theta_{13}$ 
with signs III and IV with a larger $\theta_{13}$.  We demonstrate the 
decrease of the lepton event rate with $\sin^2(2\theta_{13})$ in Fig.~4. 
Hence when the full range of current uncertainty of $\theta_{13}$ is include, 
i.e., $\sin^2(2\theta_{13})<0.1$, the sensitivity in distinguishing the MSD 
sign is lost for both baselines.

\subsection{CP violation effects}

Figures 3 and 4 show the electron event number versus the CP phase, modulo
the matter effect.  The typical total errors are also shown.  The dominant error 
is found to be statistical, i.e., from the source (i) as described at the end of 
Sec.~2.  We see that although the event rate varies significantly with the CP 
phase, as the electron event rate is not a single valued function of the CP phase, 
it is ambiguous to determine $\delta$ from the electron event number even for 
a fixed value of $\theta_{13}$.  The caveat of the uncertainty in $\theta_{13}$ 
discussed in the two previous subsections made the ambiguity even more 
serious.
  
The sensitivity of the electron event rate to the CP phase depends on the 
beam energy as shown in Fig.~5.  At some of the beam energies, e.g., 2 
and 10 GeV for $L$=2100 km and 0.7 GeV for $L$=300 km, the curves 
are quite flat, indicating a poor sensitivity to the CP phase at such beam 
energies.  Furthermore at almost no energies that one can determine a unique 
CP phase from the electron event number at either 300 km or 2100 km.
    
To investigate the sensitivity we define two ratios involving the two CP 
conserving phases: $\delta=0^\circ$ and $\delta=180^\circ$: 
\begin{eqnarray}
R^{(0^\circ)}_{\rm CP}(\delta) &\equiv& 
                              {N_e(\delta)-N_e(0^\circ) \over \Delta{N}_e(0^\circ)}, \\
R^{(180^\circ)}_{\rm CP}(\delta) &\equiv& 
                              {N_e(\delta)-N_e(180^\circ) \over \Delta{N}_e(180^\circ)}, 
\end{eqnarray}
where $N_e(\delta)$, $N_e(0^\circ)$, and $N_e(180^\circ)$ are respectively the 
electron event numbers for CP phases $\delta$, 0$^\circ$ and 180$^\circ$, and 
$\Delta N_e(0^\circ)$ and $\Delta N_e(180^\circ)$ are the total error at 
$\delta=0^\circ$ and $\delta=180^\circ$.  We can now defined the figure of
merit \cite{FoM}, i.e., the goodness of the fit, for the CP violation measurement 
as the smaller  in magnitude of the two ratios:
\begin{equation}   
F_{CP}\equiv \left[R^{(0^\circ)}_{\rm CP}(\delta),~
                           R^{(180^\circ)}_{\rm CP}(\delta)\right]_{\rm min} .
\end{equation}
In Fig.~6 we plot $F_{\rm CP}(\delta)$ versus the peak energy of the narrow 
band beam, separately for $L$=2100 and 300 km.  We show six values of
$\delta$=0$^\circ$,  30$^\circ$,   60$^\circ$,   90$^\circ$,  120$^\circ$, and
150$^\circ$.  The curves satisfy approximately the relation
$F_{CP}(180^\circ +\delta)\approx -F_{CP}(\delta)$.  Hence the curves for
$\delta=$180$^\circ$, 210$^\circ$,  240$^\circ$,  270$^\circ$, 300$^\circ$,
and 330$^\circ$ can be inferred as the negatives of the above corresponding 
curves of $\delta$ less than 180$^\circ$.  The left panel is for the 100 kt detector 
and the right panel shows the results for a 1000 kt detector.   We see that for
the 100 kt detector at both baselines the effects of the finite CP phases are within
1$\sigma$ from each other, including the CP conserving case.  If we increase the 
detector size to 1000 kt, the CP violation effects can reach to the $2\sigma$ level 
for the beams around $E_{\rm peak}\simeq$ 3-4 GeV and 6-7 GeV for 
$\delta=60^\circ$-$120^\circ$ and $240^\circ$-$300^\circ$ at L=2100 km, and 
around $E_{\rm peak} \simeq 0.7$ GeV for the similar $\delta$ ranges at 
L=300 km. 

\subsection{Effects of the uncertainty of  $\sin^2(2\theta_{13})$}

In all the above results we have used $\sin^2(2\theta_{13})=0.05$.  Since
 $\nu_\mu \to \nu_e$ is proportional to $\sin^2(2\theta_{13})$, the latter 
is a sensitive parameter for the electron event number.  Accordingly,  the 
counting experiment of the electron event number may provide a good 
measurement for the value of $\sin^2(2\theta_{13})$.

In Fig.~7 we present  the electron event number versus 
the CP phase for different $\sin^2(2\theta_{13})$ values.  The error bars
indicate the size of the estimated total errors.  From the total errors, we see 
how precisely the $\sin^2(2\theta_{13})$ value can be measured.   For example, 
for L=2100 km  the curve of $\sin^2(2\theta_{13})=0.08 (0.06)$ lies about 
$1.5\sigma$ ($3\sigma$) away from that of  $\sin^2(2\theta_{13})=0.1$. 
Then it is difficult to distinguish 0.1 from 0.08 all along the curves.  Furthermore
without knowing the CP phase, it may be difficult to distinguishing 0.1 at
one CP phase to 0.6 at another CP phase.  This ambiguity is even more 
serious for L=300 km because there is more variation of the event number 
as a function of the CP phase.        

\section{Joint analysis of baselines 2100 and 300 km}

We imagine that major efforts of the very long baseline experiments such as
H2B are the confirmation of the matter effect, the determination of the MSD 
signs, the CP phase, and $\theta_{13}$.  However, there exist difficulties
in finding unique solutions for them, given the measured electron event rate,
as demonstrated in the preceding section.  We have discussed repeatedly in 
the previous section the ambiguities caused by the current wide range of
uncertainty in $\theta_{13}$.  There are other ambiguities which are caused by 
the multi-valueness of the oscillation probability as a function of the oscillation 
parameters and the possibility of overlapping parameter regions.  To illustrate 
the latter 
ambiguity let us consider Fig.~4.  For the simplicity of argument, let us ignore 
any possible errors.  Suppose a measurement of the electron event rate is 
60 at 300 km baseline for a narrow band beam with peak energy 0.7 GeV.   
Then the CP phase can be either around 0$^\circ$ or 150$^\circ$ for 
$\sin^2(2\theta_{13})=0.05$.  Similarly, suppose a measurement at the 2100 km 
baseline gives, say, the electron event rate is 40 at 4 GeV.  Then CP phase can 
be either 150$^\circ$ or 300$^\circ$ for $\sin^2(2\theta_{13})=0.05$.  Further, 
since the value of  $\sin^2(2\theta_{13})$ is unkown, we in fact obtain a curve 
in the $\delta-\sin^2(2\theta_{13})$ plane for a given electron event number, 
as shown in Fig.~8.  Hence the measurement from only one experiment, either 
at L=300 km or at L=2100 km, is not enough to determine CP phase or the
value of $\sin^2(2\theta_{13})$. 

To illustrate the advantage of the joint analysis of two widely different baselines, 
we plot in Fig.~8  $\sin^2(2\theta_{13})$ vs $\delta$ for measured electron event 
rates for both 300 km and 2100 km baselines at respectively 60 and 40 events for
the MSD sign I.  In the absence of any errors, the intersect of the curves gives 
unique values of both $\sin^2(2\theta_{13})$ and $\delta$.  In reality the situation 
will be more complicated due to the presence of errors of the measurements, and
hence the intersect of the two curves will cover a sizable area of the 
$\sin^2(2\theta_{13})$ vs $\delta$ plane.  However, this example shows the 
possibility of extra leverages one can gain with two different baselines.   

In this section we present some of our analyses of such joint measurements,
taking the advantage of superbeams like HIPA, which can offer multiple narrow 
band beams of different energies. We use different energies at the two baselines.  
We will plot 2100 km baseline vs 300 baseline by simultaneously looking at 
two different parameters.

\subsection{$\sin^2(2\theta_{13})$ and the CP phase $\delta$}

In Fig.~9 we show electron event number at L=2100 km 
versus those at L=300 km for fix MSD sign I.  Each curve has a fixed value of 
$\sin^2(2\theta_{13})$ with the CP phase $\delta$ varies in the full possible 
range from $0^\circ$ to $360^\circ$.  The $\delta=0^\circ$ point is marked by 
a solid dot and the $\delta=180^\circ$ point by a cross.  The direction of 
increasing $\delta$ is indicated by the arrow on the curve.   
The curves are generally ellipses and the eccentricity 
of the ellipse is determined by the specific beam energies of the two baselines.  

We fix 0.7 GeV for the 300 km baseline and allow the energy at 2100 km to
change.
The upper diagram of Fig.~9 is at 4 GeV for 2100 km.  When 
$\sin^2(2\theta_{13})$ increases the ellipse moves towards the upper right, i.e., 
increasing the electron event rate for both baselines.  This is expected from
the fact that the oscillation probability $\nu_\mu\rightarrow \nu_e$ is 
proportional to $\sin^2(2\theta_{13})$.  Since the ellipses of neighboring 
values of $\sin^2(2\theta_{13})$ overlap significantly, the value of $\delta$
and $\sin^2(2\theta_{13})$ can not be determined uniquely, reflecting again
the ambiguities discussed in the preceding section.  However there are energies 
at which the overlap of the ellipses is minimized.  The lower diagram of 
Fig.~9 shows that the ellipses of constant $\theta_{13}$ are collapsed into 
lines when the beam energy 
of the 2100 km baseline is 6.3 GeV.  So in principle the joint measurement allow 
us to narrow down the allowed range of $\sin^2(2\theta_{13})$.   For the
lines each measurement still allows two values of $\delta$.  But the two 
values of $\delta$ which fall on top of one another on the line segment will 
be separated when the line becomes an ellipse.  So measurements at both 
6.3 and 4 GeV will offer a better possibility to determined the values of 
$\sin^2(2\theta_{13})$ and $\delta$ simultaneously.
 
In Table 3 we present, for the case of MSD sign I, some $E_{\rm peak}$ values 
in GeV of narrow band beams where the ellipses of Ne(300) versus Ne(2100) 
as the CP Phase varies from $0^\circ$ to $360^\circ$ collapse into lines.   At 
these energies the curves for MSD sign II are ellipses of high eccentricities
which approximate lines.  
For MSD signs III and IV, and in the absence of matter effect the
curves are ellipse of very high eccentricities.  For these energies the combined 
measurements of electron event at L=2100 km and L=300 km can provide 
better measurement for the $\sin^2(2\theta_{13})$. 

\subsection{MSD sign and the CP phase $\delta$}

In Fig.~10 we present similar results, but for different MSD signs with fixed
$\sin^2(2\theta_{13})=0.05$.  The results without the matter effect are also 
plotted, with the dotted curves denoting MSD sign II or III and 
dashed ones I or IV.
In the absence of the matter effect MSD signs I and IV 
give the same results, so do the  MSD signs II and III, as already shown in 
Tables 1 and 2.
For the almost overlapped curves of MSD sign III and IV with matter effects,
the solid ones denote III and dotted ones IV.   

It is clear from Fig.~10 that in the lower diagram, i.e., 6.3 GeV for the 2100 km
baseline, it is quite easy to differentiate MSD signs I and II from III and IV,
and from the case without the matter effect.  To make better measurements
it is again better to take measurements with the line together with the ellipse.   

\section{Conclusion}
\label{sec4}
In the above study of the event rates and the sensitivity to various  
oscillation parameters investigated, we found:  
\begin{itemize} 
{\setlength\itemsep{-0.6ex}
\item[{\rm(a)}]  At the distance L=2100 km, a narrow band beam with peak 
                energy of about 6 GeV is optimum for measuring  CP violation 
                effects and about 5 GeV for measuring matter effects.
\item[{\rm(b)}] To measure the CP violation effect at a shorter distance such 
                as L=300 km, a  narrow band beam with lower peak energy 
                ($\sim 0.7$ GeV) is 
                preferable.  But the matter effect is hardly observable at such a 
                shorter baseline.
\item[{\rm(c)}] The two baselines, 300 km and 2100 km, are complementary
                to each other.  Through the joint analysis of the two baselines, some 
                of the ambiguities associated with the measurement at either 
                baselines may be resolved.   
}
\end{itemize}
With the optimum narrow band beam,  a 5-year operation of a 100 kt water  
Cerenkov detector at a very long distance such as L=2100 km has the following  
physics prospects:
\begin{itemize} 
{\setlength\itemsep{-0.6ex}
\item[{\rm(1)}] The matter effects can be observed.
\item[{\rm(2)}] The sign of  $\Delta{\rm m}^2_{32}$ may be determined.
\item[{\rm(3)}] The sign of  $\Delta{\rm m}^2_{21}$ may be determined 
                       only in favorable situations. 
\item[{\rm(4)}] Evidence exceeding 2-$\sigma$ of a CP violating phase 
                        may be seen in favorable cases for a detector size of 
                        1000 kt or with a much longer running time.
\item[{\rm(5)}] Combined with the analyses of L=300 km, the parameter 
                $\sin^2(2\theta_{13})$ may be measured and the matter effects 
                are more clearly determined.
}
\end{itemize} 
 
In this article we have focused on the $\nu_\mu\rightarrow \nu_e$ exclusively.
The investigation of the $\tau$ appearance and the inclusion of the 
$\bar{\nu}_\mu$ beam option in the analysis, which is needed in the cases of
MSD signs III and IV, i.e., $\Delta{m}^2_{32} < 0$, will be taken for a future
investigation.  There we will also make a more complete search for the best 
energies of the two baselines for the various parameters.

We finally note that the statistics are generally low in all the cases discussed.
Running with higher energy narrow band beam will increase the statistics.
However, that may be disfavored by the figure of merit (signal to error ratio).  
Another way to increase the statistics is to increase the detector mass.  It has 
been pointed out, however, that there is a saturation problem \cite{saturation} 
caused by the systematic errors which are of the form of the errors of types 
(ii) and (iii) as discussed at the end of Sec. 2.  These errors increase linearly 
as the number of events rather than the square root of the number of events 
as is in the case of the statistical error.  Hence, when the mass of the detector
is increased so that the number of events becomes sufficiently large, the 
systematical error becomes dominant.  After that, further increase of the 
detector size may no longer be beneficial.  In Fig.~11 we show the ratio of 
$\Delta{N}_e$ to $N_e$ as a function of the detector mass.
We see that according to our general error estimate the best 
$\Delta{N}_e$ to $N_e$ ratio can be attained is 6\%.  When the detector
reaches 1000 kt the benefit of further increasing the detector size is no long
significant.  
 
\section*{Acknowledgment}
We thank K. Hagiwara and N. Okamura for discussions. We also thank our 
colleagues of the H2B collaboration \cite{H2B} for support.  This work is 
supported in part by DOE Grant No. DE-FG02-G4ER40817.


\newpage

\begin{table}
\caption{
Event rates of 5-year operation with (without) matter effects for different MSD 
sign choices for a narrow band beam of $E_{\rm peak}=4$ GeV.  The 
CP-phase is taken to be zero. }  
\label{table_1}
\vspace{7mm}
\begin{center} 
\begin{tabular}{llccc} 
\hline
\multicolumn{2}{c}{ } &electron \# &    muon \#  &    tau \# \\
            & I      & 34 (10)               & 430 (435)        & 10 (11)   \\      
L=2100 km   & II     & 46 (16)               & 405 (415)        & 11 (11)   \\ 
            & III    & 3  (16)               & 413 (415)        & 12 (11)   \\ 
            & IV     & 3  (10)               & 427 (435)        & 11 (11)  \\
            & & & & \\
            & I      &  159 (157) &          39408 (39407) &  72 (72) \\
L=300 km    & II     &  119 (116) &          39535 (39535) &  71 (71) \\ 
            & III    &  114 (116) &          39535 (39535) &  71 (71) \\ 
            & IV     &  154 (157) &          39408 (39407) &  72 (72) \\
\hline
\end{tabular}
\end{center}
\end{table}
\begin{table}
\caption{ Same as Table 1, but for a wide band beam.}
\label{table_2}
\vspace{7mm}
\begin{center} 
\begin{tabular}{llccc} 
\hline
\multicolumn{2}{c}{ } &electron \# &    muon \#  &    tau \# \\
            & I      &  151 (96)   &  2313 (2311)     &  448 (453) \\
L=2100 km   & II     &  151 (90)    & 2326 (2333)     &  443 (449) \\
            & III    &  39  (90)    & 2335 (2333)     &  454 (449) \\
            & IV     &  49  (96)    & 2308 (2311)      & 458 (453) \\
            & & & & \\
            & I      &  453  (443)  &  271536  (271535)   &   731 (731) \\
L=300 km    & II     &  359  (348)  &  271842  (271842)   &   718 (718) \\
            & III    &  337  (348)  &  271843  (271842)   &   718 (718) \\
            & IV     &  431  (443)  &  271535  (271535)   &   731 (731)  \\
\hline
\end{tabular}
\end{center}
\end{table}
\begin{table}
\caption{ Some $E_{\rm peak}$ values (GeV) of narrow band beams where the ellipses 
           of Ne(300) versus Ne(2100) as CP Phase varies from 0$^\circ$ to
           360$^\circ$ collapse into line segments. The MSD sign is assumed to be 
          case I. }
\label{table_3}
\vspace{7mm}
\begin{center} 
\begin{tabular}{c||lllll} 
\hline
$E_{\rm peak}(300)$ &  \multicolumn{5}{c}{$E_{\rm peak}(2100)$} \\ \hline                        
    0.70            &  0.750 &    1.215&  1.85 &  2.30 &   6.30 \\  \hline
    0.80            &  0.820 &    1.10 &  1.98 &  2.25 &   7.60 \\ \hline
    0.85            &  0.820 &    1.20 &  2.05 &  2.19 &   8.30 \\ \hline                 
\end{tabular}
\end{center}
\end{table}

\begin{figure}[htb]
\includegraphics[height=22cm,width=15cm,angle =0]{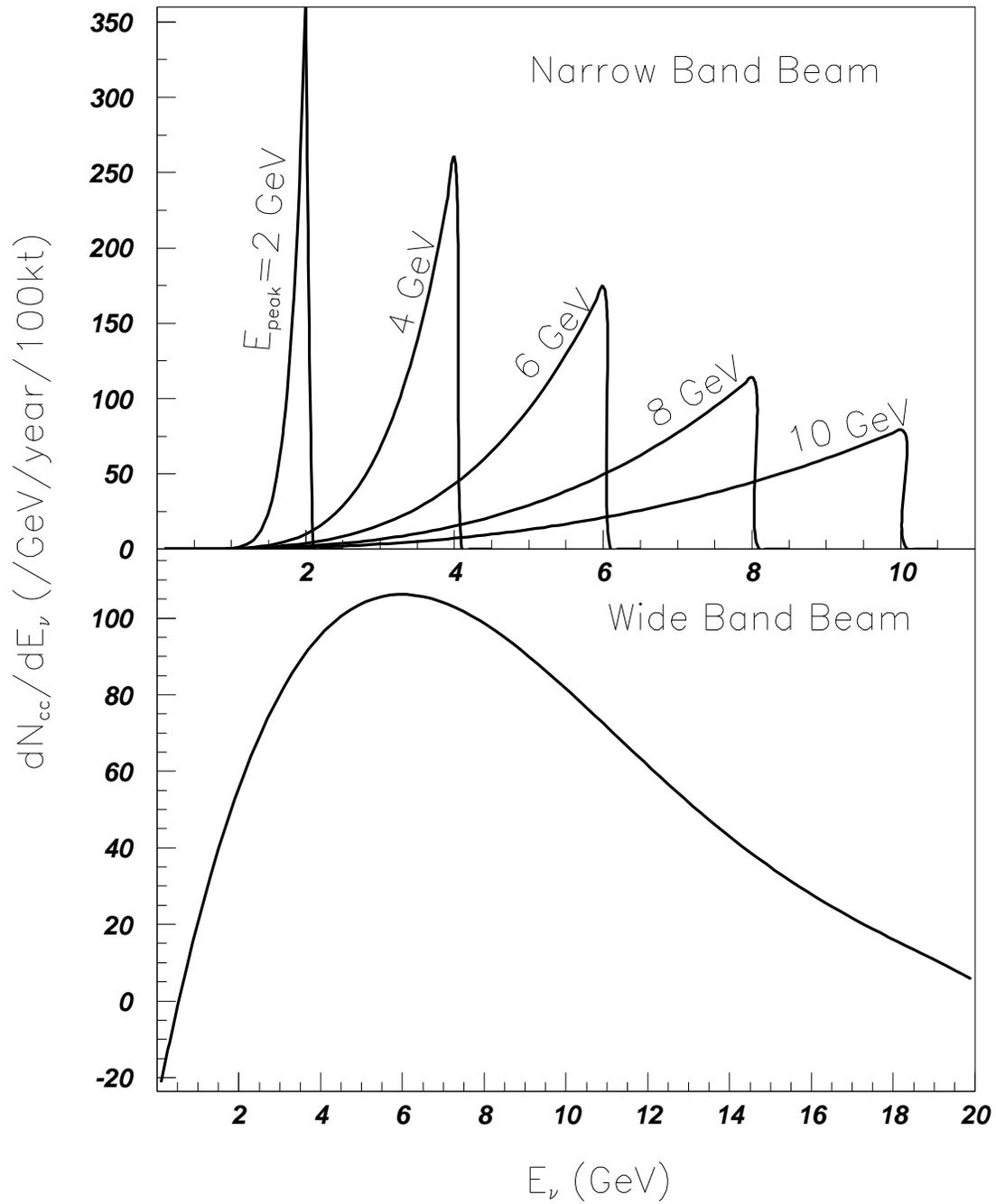}
\vspace*{-3cm}
\caption[]{ The energy $E_{\nu}$ distribution of charged-current events $N_{cc}$ 
            for one year operation of a 100 kt detector.}
\end{figure}

\begin{figure}[htb]
\hspace*{-1cm}
\includegraphics[height=22cm,width=18cm,angle =0]{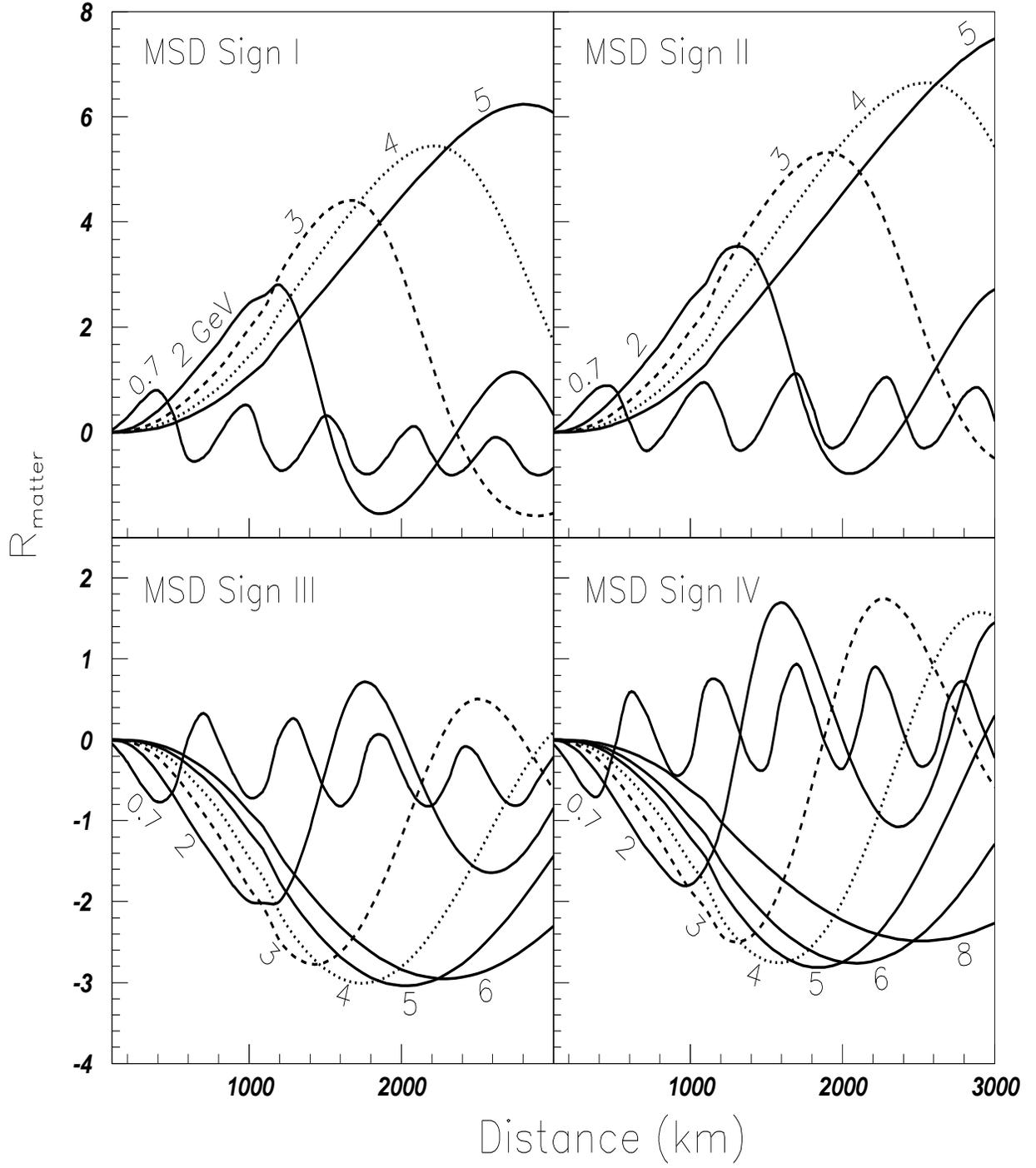}
\vspace*{-3cm}
\caption[]{  $R_{\rm matter}$ (Eq. (5)) versus the baseline for 
             several narrow band beams.  The CP phase $\delta$ is taken to be zero and
             $sin^2(2\theta_{13})=0.05$.} 
\end{figure}

\begin{figure}[htb]
\includegraphics[height=22cm,width=15cm,angle =0]{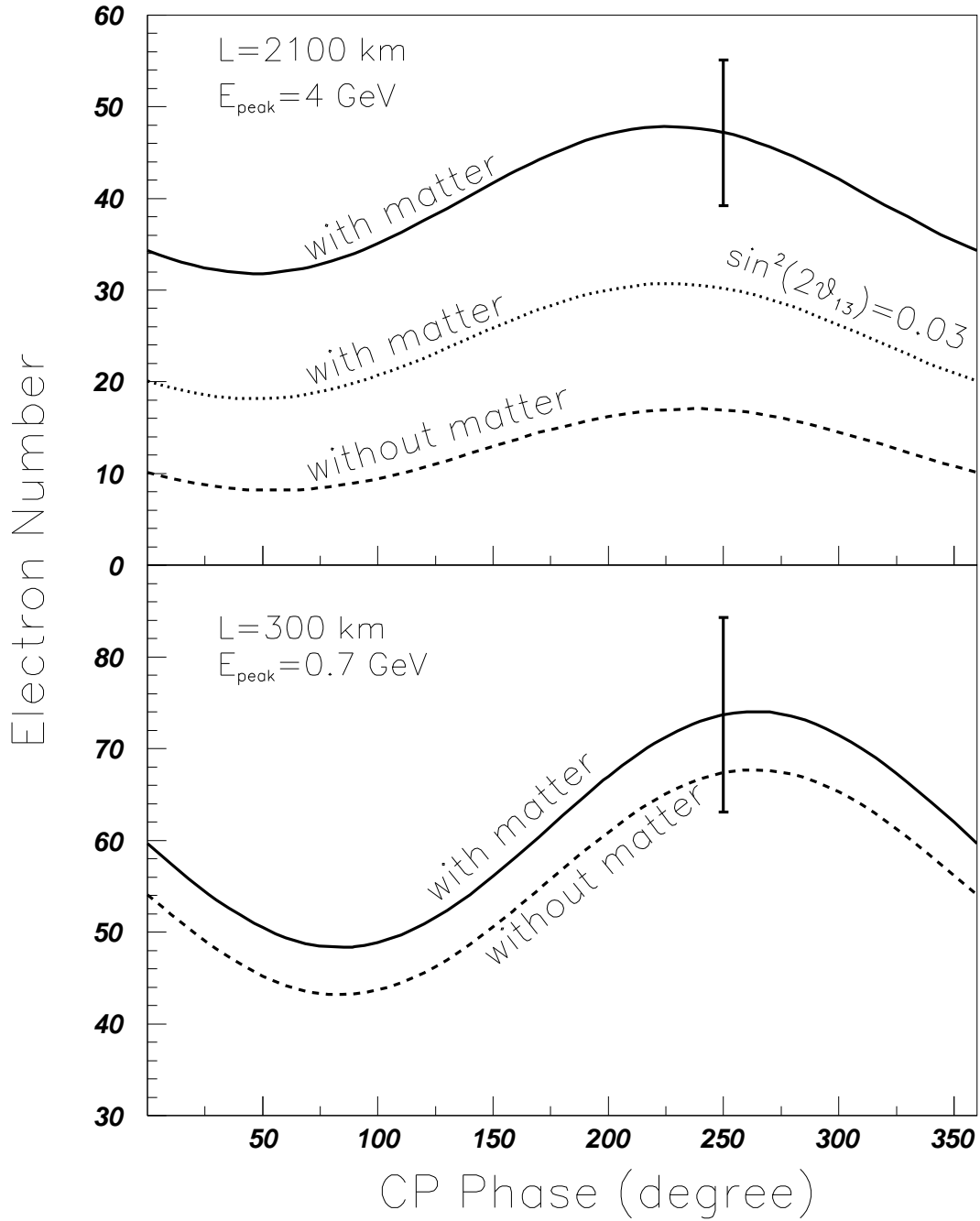}
\vspace*{-3cm}
\caption[]{ The electron event number versus the CP phase with and without 
            the matter effect.  $\sin^2(2\theta_{13})$ is assumed to be 0.05 except 
            for the dotted curve which is for $\sin^2(2\theta_{13})=0.03$ to show 
            the effect of varying $\theta_{13}$.  Representative total errors are 
            also shown.  The MSD sign is assumed to be I.} 
\end{figure}

\begin{figure}[htb]
\includegraphics[height=22cm,width=15cm,angle =0]{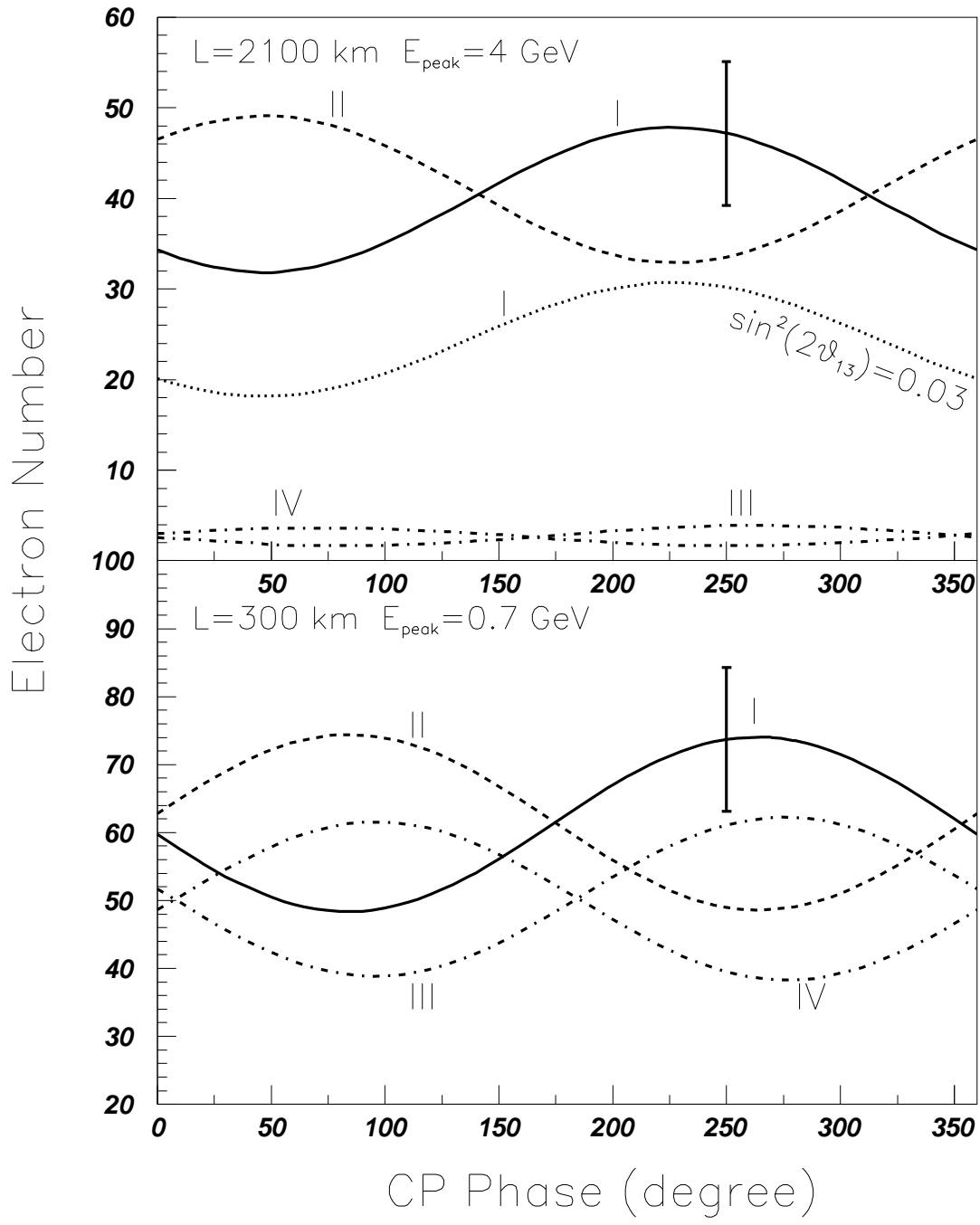}
\vspace*{-3cm}
\caption[]{ Same as Fig.~3, but for different MSD signs with  matter effect. }  
\end{figure}

\begin{figure}[htb]
\includegraphics[height=23cm,width=15cm,angle =0]{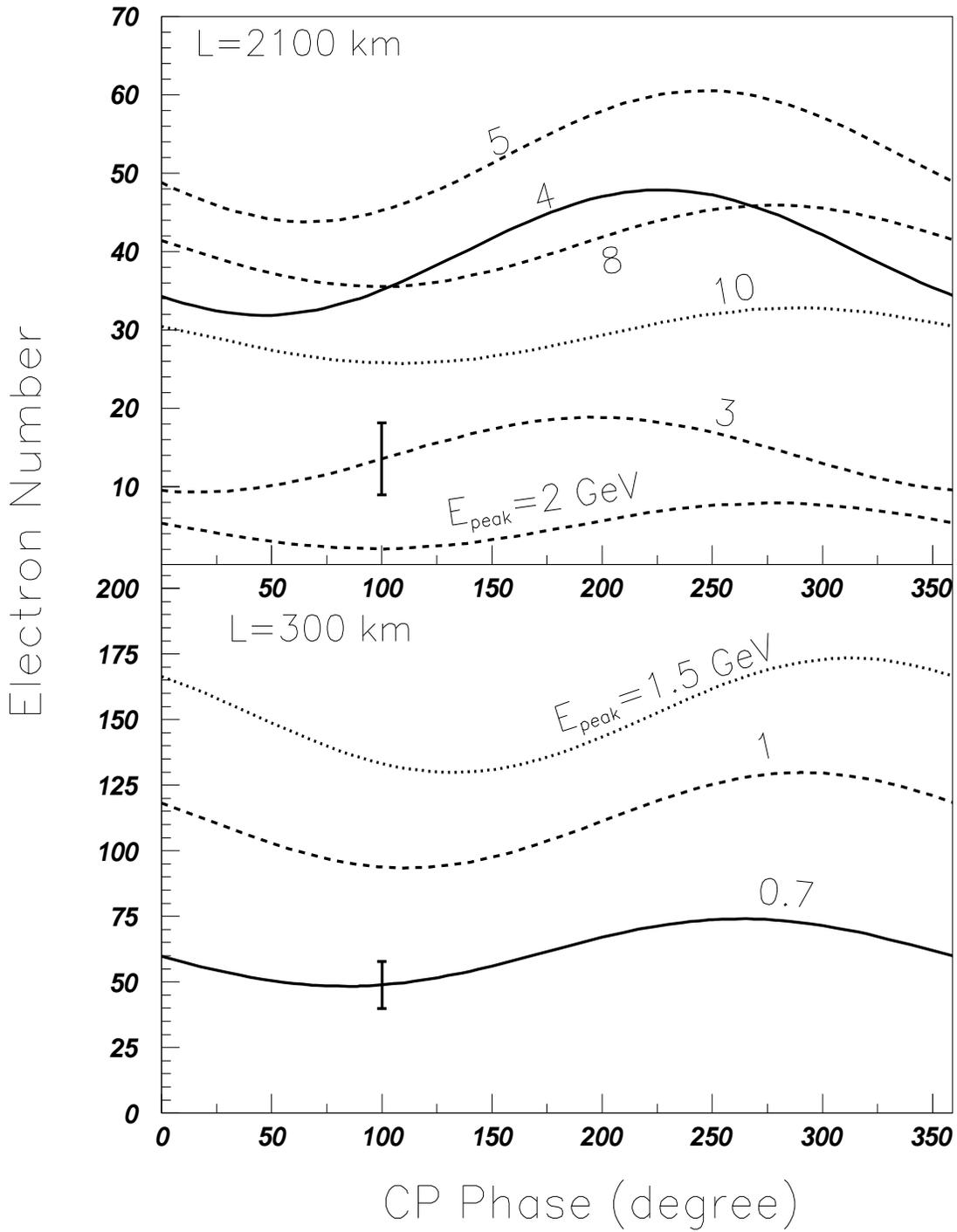}
\vspace*{-3cm}
\caption[]{  The electron event rate versus the CP phase 
             for different narrow band beams. The MSD sign is assumed to be I. }
\end{figure}
 
\begin{figure}[htb]
\hspace*{-1cm}
\includegraphics[height=22cm,width=18cm,angle =0]{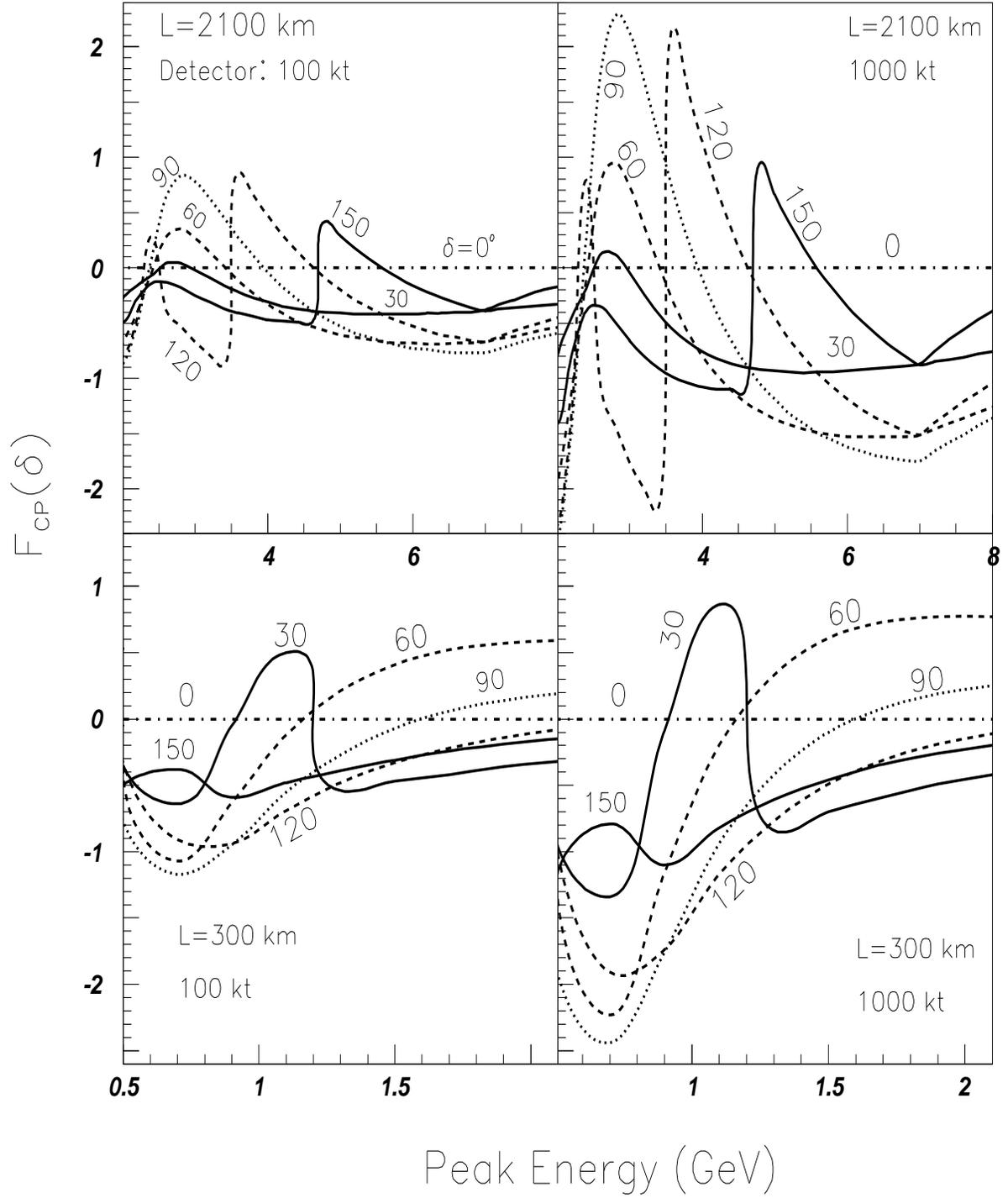}
\vspace*{-2cm}
\caption[]{$F_{CP}(\delta)$ (Eq. (8)) versus the peak energy of the narrow 
           band beams.  
           The MSD sign is assumed to be I. With the approximate relation,
           $F_{CP}(180^\circ +\delta)=-F_{CP}(\delta)$, the curves for 
           $\delta=180^\circ$, $210^\circ$, $240^\circ$, $270^\circ$, $300^\circ$,
           and $330^\circ$ can be inferred.}
\end{figure}

\begin{figure}[htb]
\includegraphics[height=22cm,width=15cm,angle =0]{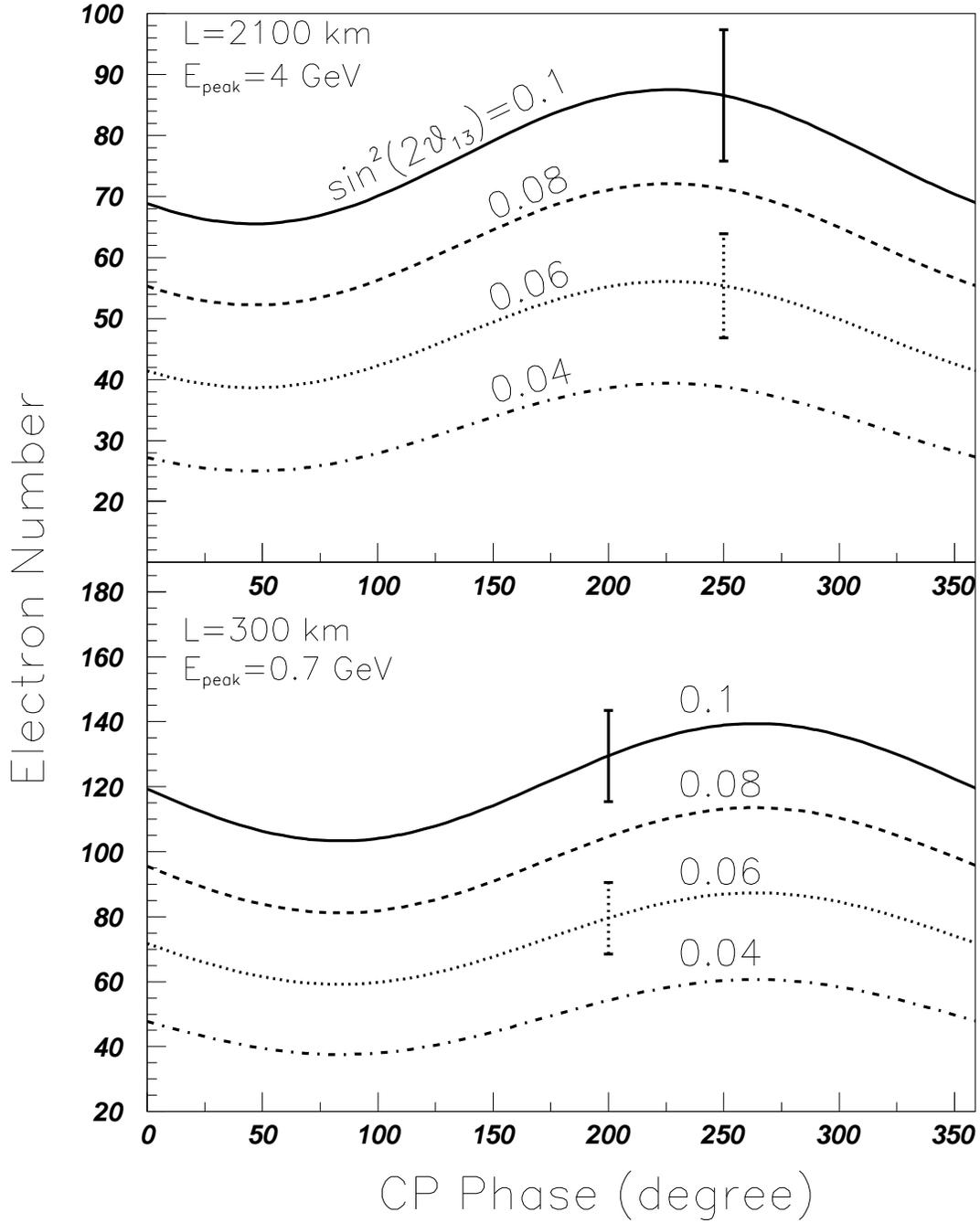}
\vspace*{-3cm}
\caption[]{ The electron event number versus the CP phase for 
            different $\sin^2(2\theta_{13})$ values.  The MSD sign is assumed to be I. 
            Total errors at some points are  also shown. }  
\end{figure}

\begin{figure}[htb]
\includegraphics[height=16cm,width=15cm,angle =0]{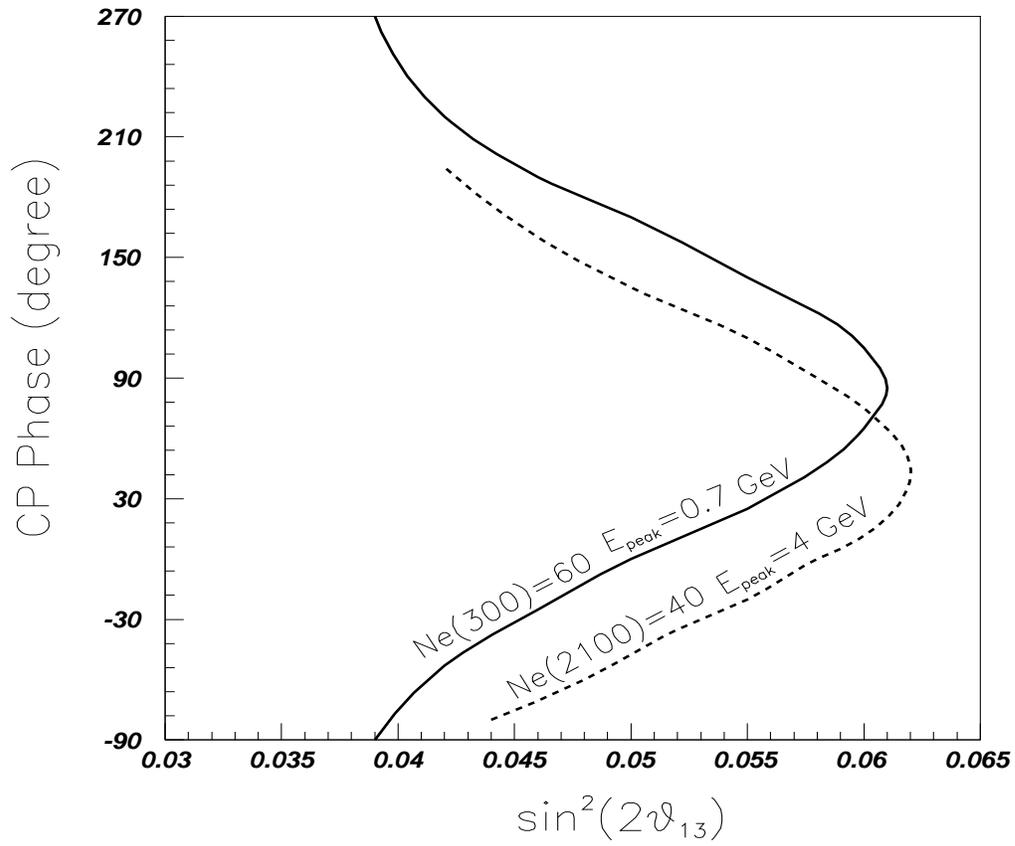}
\vspace*{-1cm}
\caption[]{ The CP phase versus  $\sin^2(2\theta_{13})$ for a given electron event number $N_e$.
            The solid (dashed) curve is for $N_e=60$ ($40$) at L=300 km (2100 km) with a narrow
            band beam $E_{\rm peak}=0.7$ GeV (4 GeV). The MSD sign is assumed to be I. }  
\end{figure}

\begin{figure}[htb]
\includegraphics[height=22cm,width=15cm,angle =0]{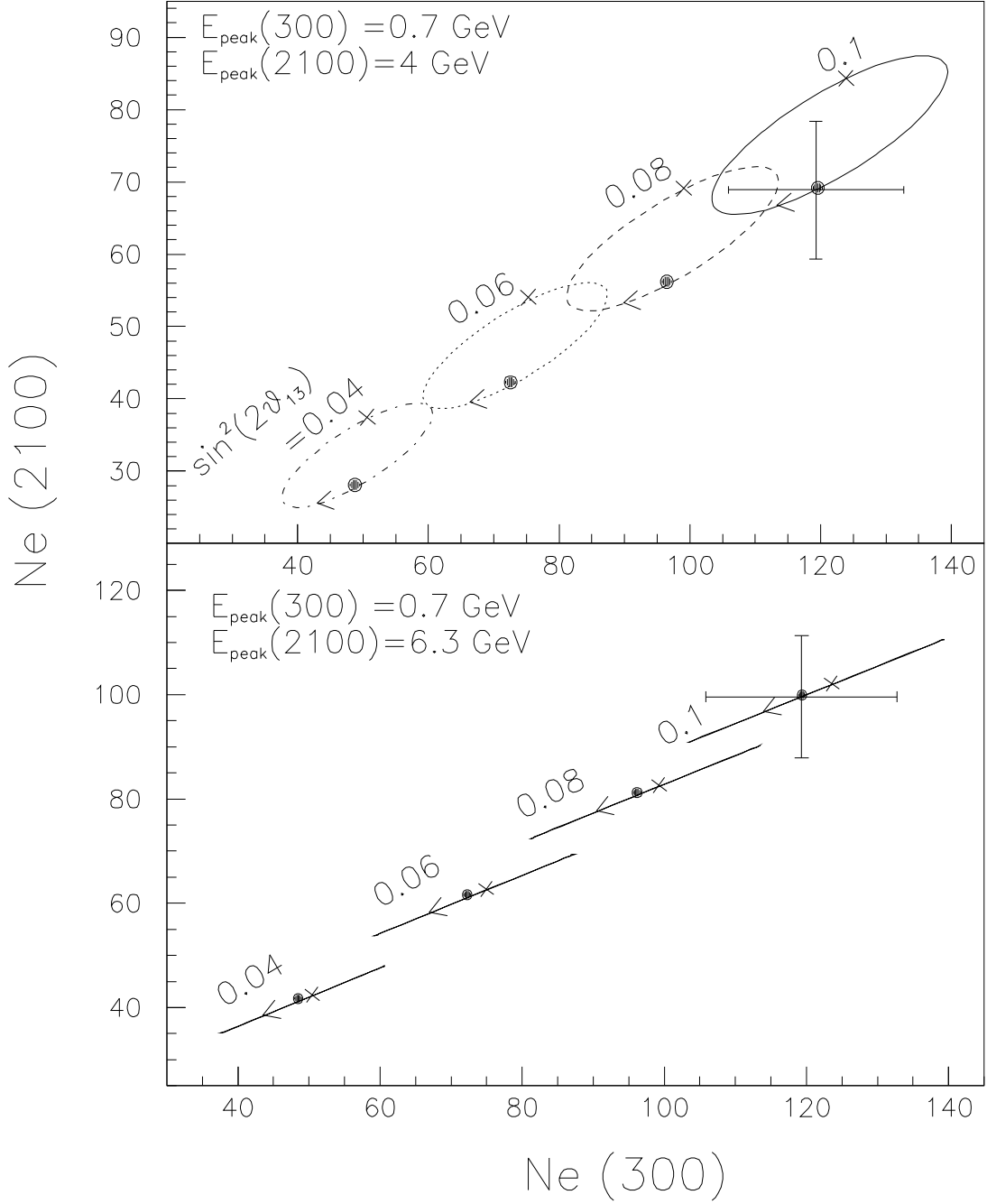}
\vspace*{-3cm}
\caption[]{\small Electron event number at L=2100 km versus  
           L=300 km for different $\sin^2(2\theta_{13})$ values. The CP phase 
           $\delta$ increases from $0^\circ$ (solid bullets) to $180^\circ$ (crosses) 
           then to $360^\circ$ according to the direction indicated by the arrows.  
          The MSD is assumed to have the sign I. In the lower digram for 
           $E_{\rm peak}(300)=0.7$ GeV and $E_{\rm peak}(2100)=6.3$ GeV, 
           the ellipses collapse into line segments. The typical total errors are
           also shown. }             
\end{figure}

\begin{figure}[htb]
\includegraphics[height=22cm,width=15cm,angle =0]{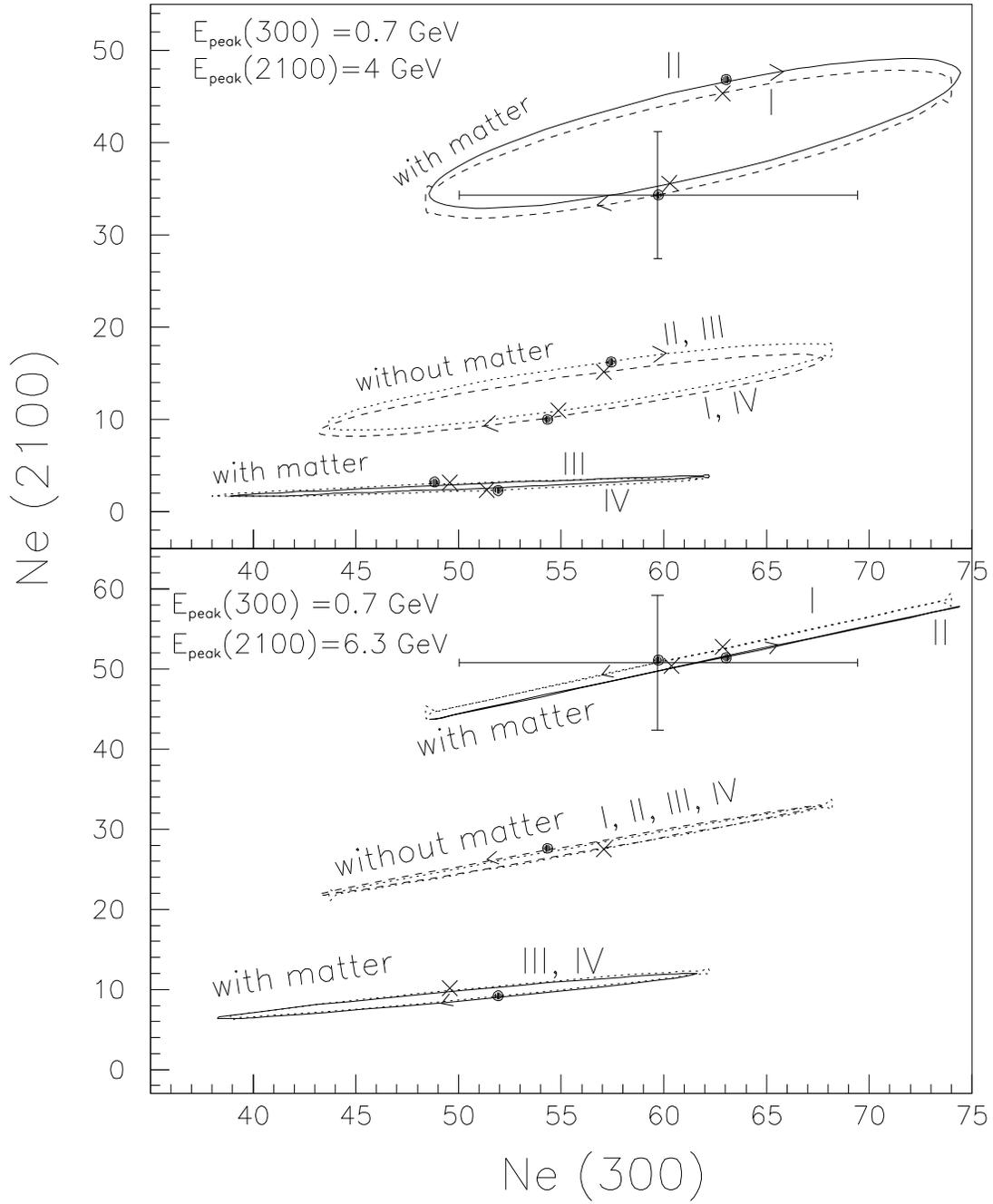}
\vspace*{-3cm}
\caption[]{Similar to Fig.~9 for different MSD signs with fixed
             $\sin^2(2\theta_{13})=0.05$. 
             The results without matter effects are also plotted.}
\end{figure}

\begin{figure}[htb]
\includegraphics[height=16cm,width=15cm,angle =0]{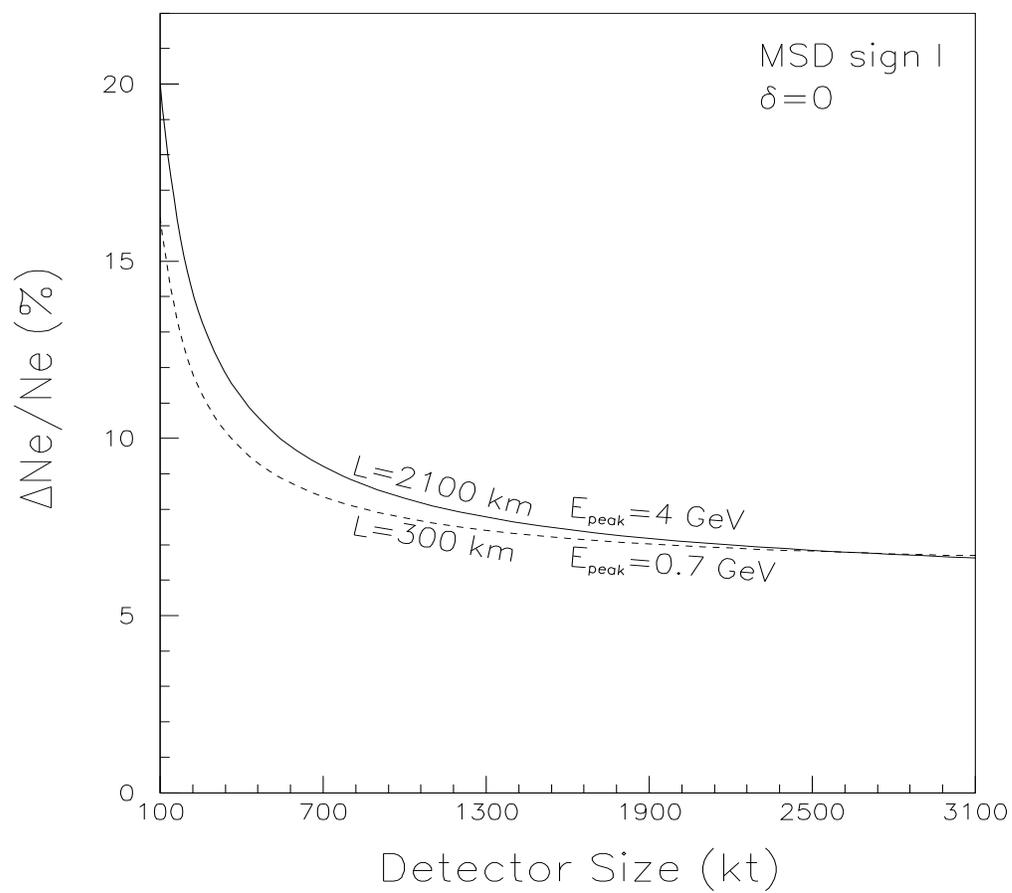}
\caption[]{ The relative error $\Delta N_e/N_e$ versus the detector size
                  for a 4 GeV narrow band beam.}
\end{figure}

\end{document}